 \documentstyle[12pt]{article}
 
\begin{document}

 \rightline{EFI 94-64}

 \medskip \begin{center} \large {\bf Dirac Monopoles and Witten's Monopole
 Equations}

 \medskip\normalsize
                Peter G.O. Freund\\

             {\em Enrico Fermi Institute and Department of Physics\\
             The University of Chicago, Chicago, IL 60637, USA}

 \end{center}
 \medskip

\noindent{\bf Abstract}: A simple solution of Witten's monopole equations
is given.

\bigskip

Witten \cite{WI} has developed a new, elegant and much simplified approach to
the Donaldson theory \cite{DO} of 4-manifolds. Instead of studying instantons
of non-abelian gauge theory, Witten's dual approach is abelian
and starts from the monopole equations

$$
F_{AB}=\frac{i}{2}(M_{A}\overline{M}_{B} + M_{B}\overline{M}_{A})
{}~ ~ ~ ~ \eqno(1a)
$$

$$
D_{A\dot{A}}M_{A}=0, ~ ~ ~ D_{A\dot{A}}=\partial_{A\dot{A}} + iA_{A\dot{A}}
{}~ ~ ~ ~ \eqno(1b)
$$

Here both dotted and undotted spinor indices take the values $1,2$ and are
raised and lowered with the corresponding invariant Levi-Civit\`{a} tensors.
$F_{AB}$ are connected with the abelian field strengths $F_{\mu\nu}$ by
$$
F_{AB}=\frac{1}{4}(\sigma^{\mu\nu})_{AB}F_{\mu\nu}, ~ ~ ~ ~
F_{\mu\nu}=\partial_{\mu}A_{\nu}-\partial_{\nu}A_{\mu}    ~ ~ ~ ~      \eqno(2)
$$

Witten noted that in flat space the equations (1) admit no $L^2$ solutions. We
wish to point out that they nevertheless admit the following very simple
{\em non}-$L^2$ solution
$$
A_0=0, ~ ~ ~ ~  \vec{A}=\frac{(-y,x,0)}{2r(r-z)}
{}~ ~ ~ ~  \eqno(3a)
$$
$$
\left( \begin{array}{c}
        M_1 \\ M_2
 \end{array} \right)=\frac{1}{2r\sqrt{r(r-z)}} \left( \begin{array}{c}
                                              x-iy \\ r-z
                                             \end{array} \right),
{}~ ~ ~ ~         \eqno(3b)
$$
so that
$$
\left( \begin{array}{c}
        \overline{M}_1 \\ \overline{M}_2
       \end{array} \right)=\frac{1}{2r\sqrt{r(r-z)}} \left( \begin{array}{c}
                                                          -(r-z) \\ x+iy
                                                        \end{array} \right).
{}~ ~ ~ ~          \eqno(3c)
$$

It is straightforward to check that the fields (3) satisfy Witten's monopole
equations (1) (we recall that for $m,n=1,2,3, ~
(\sigma^{mn})_{AB}=\epsilon^{mnp}\sigma_{p}\sigma_2$).

Notice that the potentials (3a) correspond to a singular Dirac (not to an
't Hooft-Polyakov) monopole of minimal magnetic charge. We chose to write
these potentials with Dirac-string ``showing'', rather than using sections
\cite{YW}. The spinor (3b) is also a familiar object. In fact, by
construction, the $SU(2)$ gauge transformation

$$
       \Omega= \sqrt{2}r \left( \begin{array}{cc}
                     M_1 & \overline{M}_1   \\
                     M_2 & \overline{M}_2
                    \end{array}       \right),
{}~ ~ ~ ~     \eqno(4)
$$
(with the $M$'s given by eqs. (3b) and (3c)), is such that
$M=\frac{1}{\sqrt{2}r}\Omega U$, $U$ being the unit spinor $U^{T}=(1~0)$, so
that this gauge transformation rotates \cite{AFG} a constant $SU(2)$-triplet
Higgs field into a hedgehog. The spinor $M$ thus has the property \cite{GU}
that $M^{\dag}\vec{\sigma}M$ is
a Coulomb field, which is essentially the first Witten equation (1a) for the
solution (3). What has apparently not been appreciated previously,
is the fact that
this spinor also obeys the the Weyl-Dirac equation $\sigma^{\mu}D_{\mu}M=0$
which is essentially the second Witten equation (1b) for the solution (3).

The simplicity of the solution (3) (non-$L^2$ though it is), makes it
worthwhile to explore the solubility of Witten's equations in other cases.

\medskip
I wish to thank J.Gauntlett, J. Harvey and E. Witten for useful comments.

\medskip
\footnotesize
This work was supported in part by the grant NSF: PHY-91-23780

\end{document}